\documentclass[conference,letterpaper]{IEEEtran}
\IEEEoverridecommandlockouts

\usepackage[utf8]{inputenc} 
\usepackage[T1]{fontenc}
\usepackage{ifthen}
\usepackage[cmex10]{amsmath} 

\usepackage{cite}
\usepackage{amsmath,amssymb,amsfonts}
\usepackage{algorithmic}
\usepackage{graphicx}
\usepackage{textcomp}
\usepackage{xcolor}
\usepackage{bm}
\usepackage{courier}
\usepackage{subfigure}
\newtheorem{theorem}{Theorem}

\def\BibTeX{{\rm B\kern-.05em{\sc i\kern-.025em b}\kern-.08em
    T\kern-.1667em\lower.7ex\hbox{E}\kern-.125emX}}
\begin{document}

\title{Generalized Nearest Neighbor Decoding for MIMO Channels with Imperfect Channel State Information
}

\author{\IEEEauthorblockN{Shuqin Pang and Wenyi Zhang}
	
\IEEEauthorblockA{Department of Electronic Engineering and Information Science \\
University of Science and Technology of China\\
Emails: \texttt{shuqinpa@mail.ustc.edu.cn, wenyizha@ustc.edu.cn}}
}

\maketitle

\begin{abstract}
Information transmission over a multiple-input-multiple-output (MIMO) fading channel with imperfect channel state information (CSI) is investigated, under a new receiver architecture which combines the recently proposed generalized nearest neighbor decoding rule (GNNDR) and a successive procedure in the spirit of successive interference cancellation (SIC). Recognizing that the channel input-output relationship is a nonlinear mapping under imperfect CSI, the GNNDR is capable of extracting the information embedded in the joint observation of channel output and imperfect CSI more efficiently than the conventional linear scheme, as revealed by our achievable rate analysis via generalized mutual information (GMI). Numerical results indicate that the proposed scheme achieves performance close to the channel capacity with perfect CSI, and significantly outperforms the conventional pilot-assisted scheme, which first estimates the CSI and then uses the estimated CSI as the true one for coherent decoding.
\end{abstract}

\section{Introduction}
\label{sec:intro}

Communication theory and techniques over multiple-input-multiple-output (MIMO) fading channels have been extensively studied and widely applied through the years, serving as a key driving force for improving the physical-layer performance \cite{Telatar} \cite{TSE}. There, an issue that still remains critical to date is the efficient exploitation of channel state information (CSI), which is certainly imperfect in practical systems.

With imperfect CSI, which is typically supplied to the receiver via transmitting a known pilot signal, the estimated channel state (namely, the fading matrix) is not exactly the true one, and the implementation of optimal capacity-achieving decoders (such as maximum likelihood decoder) is often prohibitively complicated. Attributed to its simplicity and robustness, the nearest neighbor decoding rule (NNDR), which selects the codeword that is the closest, in an Euclidean distance sense, to the channel output, has attracted considerable attention. With perfect CSI, the NNDR coincides with the capacity-achieving maximum likelihood decoder; while with imperfect CSI, it is generally a mismatched decoder (see, e.g., \cite{lapi-nara98} \cite{Ganti} \cite{Lapidoth} and references therein). Through the information-theoretic tool of generalized mutual information (GMI), the behavior of the NNDR for MIMO  fading channels has been studied in \cite{Weingarten} in terms of its ergodic achievable rate, and in \cite{Asyhari12} in terms of its outage probability.

Recently, a variation of the NNDR, termed the generalized nearest neighbor decoding rule (GNNDR), has been developed for scalar-input channels \cite{wang20}. The basic idea is that performance improvement is available via appropriately processing the channel output and scaling the channel input, with the aid of the CSI, before the nearest neighbor codeword search. Through analyzing the GMI, it is demonstrated that the GNNDR can exhibit evident performance gain for channels with general nonlinear input-output relationship, including fading channels with imperfect CSI.

In this paper, we extend the idea of GNNDR to MIMO fading channels. We propose a new receiver architecture called successive GNNDR (S-GNNDR), which combines the GNNDR and a successive procedure in the spirit of successive interference cancellation (SIC), similar to the well-known vertical Bell Laboratories layered space-time (V-BLAST) architecture \cite{VBLAST}. Recall that in V-BLAST, spatial multiplexing is realized by successively estimating (via a linear minimum mean squared error (MMSE) estimator), decoding, and subtracting the substreams \cite{TSE}. Here in S-GNNDR, the estimating step and the decoding step are replaced by the GNNDR, which consists of a processing function and a scaling function to play a similar but more effective role of the linear MMSE estimator in V-BLAST. Furthermore, due to the nonlinear input-output relationship in channels with imperfect CSI, the SIC step is no longer a simple subtraction but is integrated into the GNNDR when proceeding to the next substream.

As in \cite{Lapidoth} \cite{Weingarten} \cite{zhang12} \cite{wang20}, we adopt the GMI as performance measure to characterize the behavior of S-GNNDR, under independent and identically distributed (i.i.d.) Gaussian codebook ensemble. GMI is a lower bound of the mismatched capacity \cite{lapi-nara98} \cite{Ganti}, and in fact, is the largest achievable rate to ensure that the decoding error probability, averaged over the specified i.i.d. codebook ensemble, asymptotically vanishes as the coding block length grows without bound \cite{Lapidoth}. We derive the GMIs of S-GNNDR under optimal, and under a restricted linear form of, processing functions and scaling functions.

Under the ideal scenario where perfect CSI is available at the receiver, we show that the S-GNNDR is equivalent to the capacity-achieving V-BLAST. When the CSI is imperfect as supplied by a pilot signal, the S-GNNDR adopts a drastically different approach compared with the conventional pilot-assisted scheme which first estimates the CSI and uses the estimated CSI as true one for coherent decoding \cite{how much} \cite{Lapidoth} \cite{Weingarten}. Instead, empowered by the optimized processing function and scaling function, the S-GNNDR exhibits significant rate gain over the conventional linear scheme, and even gets close to the channel capacity with perfect CSI.
\begin{figure*}[h]
	\centering	
	\includegraphics[width=0.74\textwidth]{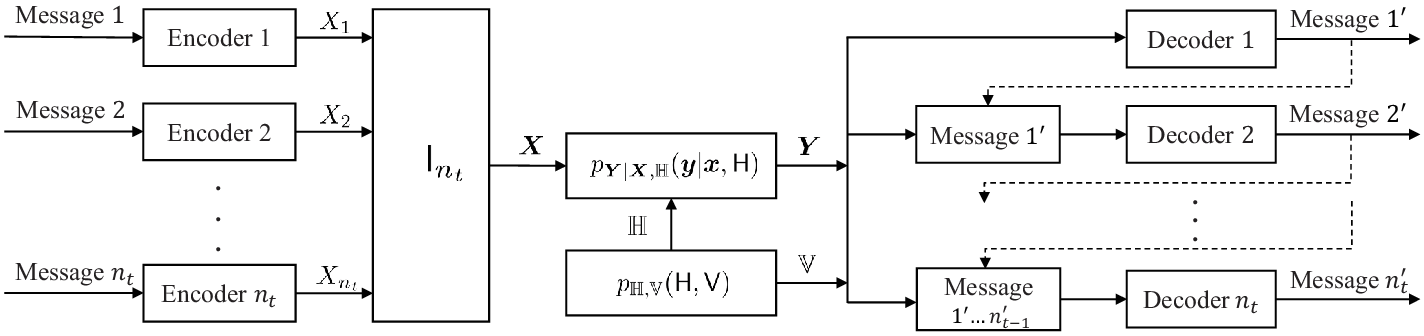}		
	\caption{Illustration of system model.}
	\label{fig:system model}
\end{figure*}

\textit{Notation:} We use uppercase letters, e.g., $X$ and lowercase letters, e.g., $x$ to denote random scalars and their realizations, respectively;  boldface capitals, e.g., $\bm{X}$ for random vectors and bold lowercase, e.g., $\bm{x}$ for their realizations; other fonts, e.g., $\mathbb{H}$ and $\mathsf{H}$ for random matrices and their realizations,  respectively.  We use superscripts $T$ and  $*$ to denote  transpose and conjugate  transpose, respectively.
\section{System Model}
\label{sec:system model}

We start with a discrete-time memoryless state-dependent channel as depicted in Figure \ref{fig:system model}, with input $\bm{X} \in \mathcal{X} \subseteq \mathbb{C}^{n_t}$, output $\bm{Y} \in \mathcal{Y} \subseteq \mathbb{C}^{n_r}$, channel state  $\mathbb{H}$ $\in \mathcal{H} \subseteq \mathbb{C}^{n_r \times n_t}$, and receiver CSI $\mathbb{V} \in \mathcal{V}$ which is assumed to belong to a general alphabet.

The channel state $\mathbb{H}$, which is the fading matrix, is independent of the input, and we consider a sufficiently long coding block length so as to render $\mathbb{H}$ to be independent over time, possibly via an ideal interleaver. Over a coding block of length $N$, we have
\begin{align}
p_{\mathbb{H},\mathbb{V}}(\mathsf{H}^N,\mathsf{V}^N)&=\prod_{n=1}^Np_{\mathbb{H},\mathbb{V}}(\mathsf{H}_n,\mathsf{V}_n),\\
p_{\bm{Y}|\bm{X},\mathbb{H}}(\bm{y}^N|\bm{x}^N,\mathsf{H}^N)&=\prod_{n=1}^Np_{\bm{Y}|\bm{X},\mathbb{H}}(\bm{y}_n|\bm{x}_n,\mathsf{H}_n).
\end{align}
We suppose that $\mathbb{V} \leftrightarrow \mathbb{H} \leftrightarrow (\bm{X}, \bm{Y})$ form a Markov chain at each channel use, meaning that the receiver CSI $\mathbb{V}$ is obtained via some mechanism independent of the current channel use. In Section \ref{sec:case studies}, we examine the scenario where $\mathbb{V}$ is in the form of a received pilot signal.

We employ the S-GNNDR architecture which is similar to V-BLAST. At the transmitter, a message stream is demultiplexed into $n_t$ substreams, separately encoded and fed to their respective transmit antennas, thus comprising the channel input $\bm{X}$.

In this paper, we adopt the i.i.d. Gaussian random codebook ensemble.
For the $i$th message substream, given a code rate $R_i$ (nats/channel use), the encoder maps a uniformly randomly selected message $m_i$ from the message set $\mathcal{M}_i=\left\{1, \cdots, \lceil e^{NR_i}\rceil\right\}$ to a length-$N$ codeword $x_i^N(m_i)$, which obeys $\mathcal{CN}(\bm{0}, P_i \mathsf{I}_N)$. The input $\bm{X}$ has a total average power constraint,  $\mathbf{E}[\bm{X}^*\bm{X}] = P$.

At the receiver, we decode the $n_t$ substreams successively. For the $i$th substream, we perform GNNDR conditioned upon the previously decoded $(i - 1)$ substreams $\bm{x}^{i-1}_n=[x_{1,n}, \cdots, x_{i-1,n}]^T$, $n = 1,\ldots,N$, in a successive fashion, which is given by
\begin{equation}
\hat{m_i}\!\!=\!\!\arg\!\!\! \min_{m_i\in \mathcal{M}_i}\!\!\sum_{n=1}^N\!\left|g_i\!\left(\bm{y}_n,\!\mathsf{V}_n,\!\bm{x}^{i-1}_n\right)\!\!-\!\!f_i\!\left(\bm{y}_n,\!\mathsf{V}_n,\!\bm{x}^{i-1}_n\right)\!x_{i,n}(m_i)\right|^2\!,
\label{eqn:GNNDR-SIC}
\end{equation}
where the mappings $g_i$ and $f_i$ are called the processing function and the scaling function, respectively \cite{wang20}. Note that in addition to the channel output $\mathbf{y}$ and the receiver CSI $\mathsf{V}$, now $g_i$ and $f_i$ also take into account the extra information provided by the previously decoded $(i - 1)$ substreams. We also remark that when $n_t=1$, the S-GNNDR architecture degenerates into  the GNNDR already established in \cite{wang20}.

\section{GMI for S-GNNDR}
\label{sec:gmi for GNND-SIC}

In this section, we employ the performance measure of GMI to characterize the behavior of S-GNNDR. As remarked in the introduction, the GMI is an achievable rate for a specified i.i.d. random codebook ensemble and a specified decoding metric, developed in the study of mismatched decoding. Indeed, it is the largest achievable rate to ensure that the decoding error probability, averaged over the specified i.i.d. random codebook ensemble, asymptotically vanishes as the coding block length grows without bound. Here, the optimized S-GNNDR and the corresponding GMI are given by the following theorem.

\begin{theorem}
\label{thm:mimo-gmi-opt}
For the information transmission system model in Figure \ref{fig:system model}, under the S-GNNDR of \eqref{eqn:GNNDR-SIC}, the maximum GMI achieved by the $i$th substream is
\begin{equation}
I^{(i)}_{\mathrm{GMI}, \mathrm{opt}} = \mathbf{E}\left[\log\frac{P_i}{w\left(\bm{Y},\mathbb{V},\bm{X}^{i-1}\right)}\right],
\label{eqn:thm1-gmi-opt}
\end{equation}
where 
\begin{equation}
\begin{aligned}
w\left(\bm{y},\!\mathsf{V},\!\bm{x}^{i-1}\right)\!\!&=\!\mathbf{E}\!\left[|X_i|^2|\bm{y},\!\mathsf{V},\!\bm{x}^{i-1}\right]-\left|\mathbf{E}\!\left[X_i|\bm{y},\!\mathsf{V},\!\bm{x}^{i-1}\right]\right|^2\\
&=\mathbf{var}\!\left[X_i|\bm{y},\!\mathsf{V},\!\bm{x}^{i-1}\right],
\end{aligned}
\end{equation}
and the corresponding optimized GNNDR is 
\begin{equation}
\begin{aligned}
&\hat{m_i}\!\!=\!\arg \!\!\!\min_{m_i\in\mathcal{M}_i}
\!\sum_{n=1}^N\!\frac{1}{\left[P_i\!-w(\bm{y}_n,\!\mathsf{V}_n,\!\bm{x}^{i-1}_n)\right]w(\bm{y}_n,\!\mathsf{V}_n,\!\bm{x}^{i-1}_n)}\\
&\cdot\left|\mathbf{E}[X_i|\bm{y}_n,\!\mathsf{V}_n,\!\bm{x}^{i-1}_n]-\frac{P_i-w(\bm{y}_n,\!\mathsf{V}_n,\!\bm{x}^{i-1}_n)}{P_i}x_{i,n}(m_i)\right|^2.
\label{eqn:thm1-gnndr}
\end{aligned}
\end{equation}

The sum GMI over the $n_t$ substreams together is
\begin{equation}\label{eqn:sum-gmi-optimal}
I^{\mathrm{sum}}_{\mathrm{GMI},\mathrm{opt}}=I^{(1)}_{\mathrm{GMI},\mathrm{opt}}+\cdots+I^{(n_t)}_{\mathrm{GMI},\mathrm{opt}}.
\end{equation}
\end{theorem}

{\it Proof:} This theorem is a consequence of \cite[Thm. 1]{wang20}, and here we provide an outline of its proof (see also \cite{Lapidoth}).

The decoding metric for \eqref{eqn:GNNDR-SIC} is defined as
\begin{equation}
D_i(m_i) = \frac{1}{N}\sum_{n=1}^N\!\left|g_i\!\left(\bm{y}_n,\!\mathsf{V}_n,\!\bm{x}^{i-1}_n\right)\!\!-\!\!f_i\!\left(\bm{y}_n,\!\mathsf{V}_n,\!\bm{x}^{i-1}_n\right)\!x_{i,n}(m_i)\right|^2\!.
\label{eqn:metric}
\end{equation}
Due to symmetry, the decoding error probability averaged over the i.i.d. Gaussian codebook ensemble is identical for all messages. Consequently, without loss of generality, it suffices to assume that the message $m_i = 1$ is transmitted. For $m_i = 1$, we have
\begin{equation}
\begin{aligned}
\lim_{N\to \infty}\! D_i(1)\!=\!\mathbf{E}\left[\left|g_i\!\left(\bm{Y},\!\mathbb{V},\!\bm{X}^{i-1}\right)\!\!-\!\!f_i\!\left(\bm{Y},\!\mathbb{V},\!\bm{X}^{i-1}\right)\!X_i\right|^2\right]&,\\
\mbox{almost surely (a.s.)}&,
\end{aligned}
\end{equation}
according to the law of large numbers. The GMI is the asymptotic exponent of the probability that an incorrect codeword corresponding to any $m_i \neq 1$ accumulates a metric $D_i(m_i)$ no greater than $\lim_{N \rightarrow \infty} D_i(1)$, and is given by, for any fixed $g_i$ and $f_i$, 
\begin{equation}
\begin{aligned}
I^{(i)}_{\mathrm{GMI}}\!\!=\!\max_{\!\theta<0,g_i,f_i}\!\bigg\{&\!\theta \mathbf{\mathbf{E}}\!\left[ \left|g_i\!\left(\bm{Y},\!\mathbb{V},\!\bm{X}^{i-1}\right)\!\!
-\!\!f_i\!\left(\bm{Y},\!\mathbb{V},\!\bm{X}^{i-1}\right)\!X_i\right|^2\right]\\
&-\Lambda_i(\theta)\bigg\},
\end{aligned}
\end{equation}
\begin{equation}
\begin{aligned}
\Lambda_i(\theta)&=\lim_{N\to \infty}\frac{1}{N}\Lambda_{i, N}(N\theta),\\
\Lambda_{i, N}(N\theta)&=\log \mathbf{E}\left[e^{N\theta D_i(m_i)}\big|\mathbb{Y},\mathbb{V},\mathbb{X}^{i-1}\right], \forall m_i \neq 1,
\end{aligned}
\end{equation}
where  $\mathbb{Y}=\left[\bm{Y}_1,\cdots,\bm{Y}_N\right]$, $\mathbb{V}=\left[\mathbb{V}_1,\cdots,\mathbb{V}_N\right]$, $\mathbb{X}^{i-1}=[\bm{X}^{i-1}_1,\cdots,\bm{X}^{i-1}_N]$. By noting that conditioned upon $\left(\bm{Y}, \mathbb{V}, \bm{X}^{i-1}\right)$, $\left|g_i\!\left(\bm{Y},\!\mathbb{V},\!\bm{X}^{i-1}\right)\!\!
-\!\!f_i\!\left(\bm{Y},\!\mathbb{V},\!\bm{X}^{i-1}\right)\!X_i\right|^2$ obeys a non-central chi-squared distribution, following \cite{Lapidoth} \cite[App. A and C]{zhang12}, we deduce that
\begin{equation}
\begin{aligned}
I^{(i)}_{\mathrm{GMI}}\!\!=\!\!\max_{\!\theta<0,g_i,f_i}\!&\bigg\{\!\theta \mathbf{\mathbf{E}}\!\left[ \left|g_i\!\left(\bm{Y},\!\mathbb{V},\!\bm{X}^{i-1}\right)\!\!
-\!\!f_i\!\left(\bm{Y},\!\mathbb{V},\!\bm{X}^{i-1}\right)\!X_i\right|^2\right]\\
&-\mathbf{E}\left[\frac{\theta \left|g_i\left(\bm{Y},\mathbb{V},\bm{X}^{i-1}\right)\right|^2}{1-\theta\left|f_i\left(\bm{Y},\mathbb{V},\bm{X}^{i-1}\right)\right|^2P_i}\right]\\
&+\mathbf{E}\left[\log \left(1-\theta\left|f_i(\bm{Y},\mathbb{V},\bm{X}^{i-1})\right|^2P_i\right)\right]\bigg\}.
\label{eqn:gmi-i}
\end{aligned}
\end{equation}
Solving the optimization in \eqref{eqn:gmi-i} as in \cite[Thm. 1]{wang20}, we obtain \eqref{eqn:thm1-gmi-opt} and the corresponding optimized GNNDR is given by \eqref{eqn:thm1-gnndr}. The sum GMI \eqref{eqn:sum-gmi-optimal} immediately follows by summing all the $n_t$ substreams according to the procedure of S-GNNDR.
{\bf Q.E.D.}

We may also evaluate the GMI for processing functions and scaling functions of restricted form. In particular, consider the following:
\begin{equation}
\begin{aligned}
g_i\left(\bm{y},\mathsf{V},\bm{x}^{i-1}\right)&=\beta_i^*(\mathsf{V})\left(\bm{y}-\bm{\gamma}^{i-1}(\mathsf{V})\bm{x}^{i-1}\right),\\
f_i\left(\bm{y},\mathsf{V},\bm{x}^{i-1}\right)&=f_i\left(\mathsf{V}\right),
\label{eqn:g-f-lin}
\end{aligned}
\end{equation}
where $\bm{\gamma}^{i-1}(\mathsf{V})=[\gamma_1(\mathsf{V}),\cdots,\gamma_{i-1}(\mathsf{V})]$, and $\beta_i(\mathsf{V})$ and $\gamma_i(\mathsf{V})$ are column vector functions of the receiver CSI $\mathsf{V}$ only. The processing function $g_i$ is a linear function of the output after subtracting a scaled version of the already decoded input substreams. As a motivation for the form of \eqref{eqn:g-f-lin}, we will show that for MIMO fading channels with perfect CSI, this returns to the  standard V-BLAST architecture.

For \eqref{eqn:g-f-lin}, the resulting optimized GMI is given as follows.

\begin{theorem}
\label{thm:mimo-gmi-lin}
For the information transmission system model in Figure \ref{fig:system model}, under the S-GNNDR of \eqref{eqn:GNNDR-SIC} with restricted form \eqref{eqn:g-f-lin}, the resulting optimized GMI achieved by the $i$th substream is
\begin{equation}
I^{(i)}_{\mathrm{GMI},\mathrm{lin}}\!\!=\!\!\mathbf{E}\!\left[\!\log\frac{P_i}{P_i\!-\!\mathbf{E}\!\left[X_i^*\bm{Y}|\mathbb{V}\right]^*\!\mathbf{E}[\tilde{\bm{Y}}\tilde{\bm{Y}}^*\big|\mathbb{V}]^{-1}\!\mathbf{E}\left[X_i^*\bm{Y}|\mathbb{V}\right]}\right],
\label{eqn:mimo-gmi-lin}
\end{equation}
where $\tilde{\bm{Y}}=\bm{Y}-\bm{\gamma}^{i-1}(\mathbb{V})\bm{X}^{i-1}$. The GMI \eqref{eqn:mimo-gmi-lin} is achieved by letting $g_i(\bm{y},\mathsf{V},\bm{x}^{i-1})=\sqrt{Q_i(\mathsf{V})}\times\tilde{\beta}_i^*(\mathsf{V})\tilde{\bm{y}}$ and $f_i(\mathsf{V})=\sqrt{Q_i(\mathsf{V})}\times\tilde{f}_i(\mathsf{V})$, where\footnote{We assume that $\mathbf{E}[\tilde{\bm{Y}}\tilde{\bm{Y}}^*|\mathsf{V}]$ is invertible.}
\begin{equation}
\begin{aligned}
\gamma_i(\mathsf{V})&=\frac{\mathbf{E}[X_i^*\bm{Y}|\mathsf{V}]}{P_i},\;\tilde{\beta}_i(\mathsf{V})=\mathbf{E}[\tilde{\bm{Y}}\tilde{\bm{Y}}^*|\mathsf{V}]^{-1}\mathbf{E}[X_i^*\bm{Y}|\mathsf{V}],\\
\tilde{f}_i(\mathsf{V})&=\frac{\mathbf{E}[X_i^*\bm{Y}|\mathsf{V}]^*\mathbf{E}[\tilde{\bm{Y}}\tilde{\bm{Y}}^*|\mathsf{V}]^{-1}\mathbf{E}[X_i^*\bm{Y}|\mathsf{V}]}{P_i},\\
Q_i(\mathsf{V})&=\dfrac{1}{P_i^2\tilde{f}(\mathsf{V})\left[1-\tilde{f}(\mathsf{V})\right]}.
\end{aligned}
\end{equation}

The sum GMI over the $n_t$ substreams together is
\begin{equation}
I^{\mathrm{sum}}_{\mathrm{GMI},\mathrm{lin}}=I^{(1)}_{\mathrm{GMI},\mathrm{lin}}+\cdots+I^{(n_t)}_{\mathrm{GMI},\mathrm{lin}}.
\end{equation}
\end{theorem}

{\it Proof:} The proof is obtained by substituting \eqref{eqn:g-f-lin} into \eqref{eqn:metric}, and optimizing $\beta_i$, $\gamma_i$ and $f_i$, which can be solved following similar steps as in the proof of \cite[Prop. 4]{wang20}.{\bf Q.E.D.}

\section{MIMO Channels with Imperfect CSI}
\label{sec:case studies}

In this section, we apply the derived GMIs to MIMO channels with imperfect CSI, to illustrate the potential performance gain of S-GNNDR.

\subsection{MIMO Channels with Perfect CSI}

Before studying MIMO channels with imperfect CSI, let us verify that the proposed S-GNNDR is equivalent to the standard V-BLAST for MIMO channels with perfect CSI at the receiver, i.e., $\mathbb{V}=\mathbb{H}$.

Consider the MIMO fading channel with $n_t$ transmit antennas and $n_r$ receive antennas,
\begin{equation}
\begin{aligned}
\bm{Y}&=\mathbb{H}\bm{X}+\bm{Z}\\
&=\bm{H}_1X_1+\bm{H}_2X_2+\cdots+ \bm{H}_{n_t}X_{n_t}+\bm{Z},
\label{eqn:channel}
\end{aligned}
\end{equation}
where $\mathbb{H}=[\bm{H}_1,\cdots,\bm{H}_{n_t}]$,
$\bm{Z}\sim \mathcal{CN}(\bm{0},\sigma^2\mathsf{I}_{n_r})$. The input covariance matrix is $\mathsf{K}_x=\mathbf{E}[\bm{X}\bm{X}^*]=\mathrm{diag}\{P_1,\cdots, P_{n_t}\}$. Defining $\bm{Z}_k= \sum_{i=k+1}^{n_t}\bm{H}_iX_i+\bm{Z}$, we have $\bm{Z}_k|\mathbb{H}\sim \mathcal{CN}(\bm{0},\mathsf{K}_{z_k})$, where
$\mathsf{K}_{z_k}=\sum_{i=k+1}^{n_t}P_i\bm{H}_i\bm{H}_i^*+\sigma^2\mathsf{I}_{n_r}$.

Let us inspect $w\left(\bm{y},\!\mathsf{H},\!\bm{x}^{i-1}\right)\!\!=\mathbf{var}\!\left[X_i|\bm{y},\!\mathsf{H},\!\bm{x}^{i-1}\right]$ in Theorem \ref{thm:mimo-gmi-opt}. According to linear
estimation theory (see, e.g., \cite{Poor}), upon observing $(\bm{y},\mathsf{H},\bm{x}^{i-1})$, the conditional probability distribution of $X_i|\bm{y},\mathsf{H},\bm{x}^{i-1}$ is $\mathcal{CN}(\mathbf{E}\!\left[X_i|\bm{y},\!\mathsf{H},\!\bm{x}^{i-1}\right],\mathbf{var}\!\left[X_i|\bm{y},\!\mathsf{H},\!\bm{x}^{i-1}\right])$, where
\begin{equation}
\begin{aligned}
\mathbf{E}\!\left[X_i|\bm{y},\!\mathsf{H},\!\bm{x}^{i-1}\right]\!\!&=\!\frac{P_i}{1+P_i\bm{h}_i^*\mathsf{K}_{z_i}^{-1}\bm{h}_i}\!\bm{h}_i^*\mathsf{K}_{z_i}^{-1}\!\left(\!\bm{y}\!-\!\sum_{k=1}^{i-1}\bm{h}_kx_k\!\right),\\
\mathbf{var}\!\left[X_i|\bm{y},\!\mathsf{H},\!\bm{x}^{i-1}\right]\!\!&=\!\frac{P_i}{1+P_i\bm{h}_i^*\mathsf{K}_{z_i}^{-1}\bm{h}_i}.
\end{aligned}
\end{equation}

Note that here $\mathbf{var}\!\left[X_i|\bm{y},\!\mathsf{H},\!\bm{x}^{i-1}\right]$ does not depend upon $\bm{y}$ or $\bm{x}^{i-1}$. We can hence obtain from Theorems \ref{thm:mimo-gmi-opt} and \ref{thm:mimo-gmi-lin},
\begin{equation}
\begin{aligned}
I^{(i)}_{\mathrm{GMI},\mathrm{opt}}&=I^{(i)}_{\mathrm{GMI},\mathrm{lin}}=\mathbf{E}\left[\log\left(1+P_i\bm{H}_i^*\mathsf{K}_{z_i}^{-1}\bm{H}_i\right)\right],
\end{aligned}
\end{equation}
and by employing the matrix determinant lemma, we further deduce that
\begin{equation}
\begin{aligned}
I^{\mathrm{sum}}_{\mathrm{GMI},\mathrm{opt}}=I^{\mathrm{sum}}_{\mathrm{GMI},\mathrm{lin}}&=\sum_{i=1}^{n_t}\mathbf{E}\left[\log\left(1+P_i\bm{H}_i^*\mathsf{K}_{z_i}^{-1}\bm{H}_i\right)\right]\\
&=\mathbf{E}\left[\log\det\left(\mathsf{I}_{n_r}+\frac{1}{\sigma^2}\mathbb{H}\mathsf{K}_x\mathbb{H}^*\right)\right].
\label{eqn:perfect-csi}
\end{aligned}
\end{equation}

When we choose $\mathsf{K}_x=\frac{P}{n_t} \mathsf{I}_{n_t}$, \eqref{eqn:perfect-csi} can be simplified into
\begin{equation}
\begin{aligned}
I^{\mathrm{sum}}_{\mathrm{GMI,opt}}&=I^{\mathrm{sum}}_{\mathrm{GMI,lin}}
=\mathbf{E}\left[\log\det\left(\mathsf{I}_{n_r}+\frac{\mathsf{SNR}}{n_t}\mathbb{H}\mathbb{H}^*\right)\right],
\label{eqn:mimo-capacity}
\end{aligned}
\end{equation}
with $\mathsf{SNR}=\frac{P}{\sigma^2}$ as the channel signal-to-noise ratio (SNR).  When $\bm{H}_1, \bm{H}_2, \cdots, \bm{H}_{n_t}$ obey i.i.d. $\mathcal{CN}(\bm{0},\mathsf{I}_{n_r})$,  it is clear that \eqref{eqn:mimo-capacity} coincides with the capacity of the channel \eqref{eqn:channel} \cite{Telatar}.

\subsection{MIMO Channels with Imperfect CSI}

In any realistic information transmission system the CSI can never be perfect. Suppose that the imperfect CSI is in the form of a received pilot signal. In this paper, for simplicity, we assume that the channel obeys block fading,\footnote{Extensions to more general time-varying fading models can be obtained in similar fashion and are left for future study.} and in each coherence interval there is a pilot signal matrix $\mathbb{X}_{\tau}\in \mathbb{C}^{T_{\tau}\times n_t}$ sent over $T_{\tau} \geq n_t$ time instants for training. According to \cite{how much}, we set $T_{\tau} = n_t$ and $\mathbb{X}_{\tau} = x_\mathrm{p}\mathsf{I}_{n_t}$;
that is, transmitting a pilot symbol from a single transmit antenna at each time instant, so that the received pilot signal is\footnote{In subsequent analysis, we will omit the overhead of the pilot signal when calculating the achievable rate, in order to be consistent with the information transmission system model in Section \ref{sec:system model}. For a block-fading channel of coherence interval length $T$, the resulting information rate hence should be discounted by a factor of $(1-T_\tau/T)$.}
\begin{equation}
\begin{aligned}
\bm{Y}_{\mathrm{p},i}=\bm{H}_i x_\mathrm{p} + \bm{Z}_{\mathrm{p},i},\quad i=1,2,\cdots,n_t.
\end{aligned}
\end{equation}
The CSI $\mathbb{V}$ is the received pilot signal,  $\mathbb{V}=\mathbb{Y}_\mathrm{p}=[\bm{Y}_{\mathrm{p},1},\cdots,\bm{Y}_{\mathrm{p},n_t}]$.

According to Theorem \ref{thm:mimo-gmi-opt}, under S-GNNDR, the maximum GMI achieved by the $i$th substream is
\begin{equation}
I^{(i)}_{\mathrm{GMI},\mathrm{opt}}=\mathbf{E}\left[\log\frac{P_i}{w\left(\bm{Y},\mathbb{Y}_\mathrm{p},\bm{X}^{i-1}\right)}\right],
\end{equation}
and the maximum sum GMI is
\begin{equation}
	I^{\mathrm{sum}}_{\mathrm{GMI},\mathrm{opt}}=I^{(1)}_{\mathrm{GMI},\mathrm{opt}}+\cdots+I^{(n_t)}_{\mathrm{GMI},\mathrm{opt}}.
\end{equation}

Different from conventional pilot-assisted schemes where the received pilot signal is used for estimating the channel fading matrix, which is then treated as the true channel state for coherent decoding, the S-GNNDR adopts a direct approach, which bypasses the channel estimation step and directly estimates the input for decoding, based on the output and the received pilot signal.

We now turn to evaluating the GMI in Theorem \ref{thm:mimo-gmi-lin}, assuming that $\bm{H}_i\sim \mathcal{CN}\left(\bm{0},\eta_i^2\mathsf{I}_{n_r}\right)$. We will show that the resulting GMI exactly coincides with that achieved by the conventional pilot-assisted scheme; see, e.g., \cite{how much} \cite{Weingarten}.

First, we can derive that
\begin{equation}
\begin{aligned}
\mathbf{E}[X_i^*\bm{Y}|\mathsf{Y}_\mathrm{p}]&=P_i\mathbf{E}[\bm{H}_i|\mathsf{Y}_\mathrm{p}],\\
\mathbf{E}[\tilde{\bm{Y}}\tilde {\bm{Y}}^*|\mathsf{Y}_\mathrm{p}]&=\mathbf{E}[\bm{Y}\bm{Y}^*|\mathsf{Y}_\mathrm{p}]-\sum_{k=1}^{i-1}\frac{\mathbf{E}[X_k^*\bm{Y}|\mathsf{Y}_\mathrm{p}]\mathbf{E}[X_k^*\bm{Y}|\mathsf{Y}_\mathrm{p}]^*}{P_k}.
\end{aligned}
\end{equation}
According to linear estimation theory (see, e.g., \cite{Poor}), we have
\begin{equation}
\begin{aligned}
\mathbf{E}[\bm{H}_i|\mathsf{Y}_\mathrm{p}]&=\frac{\eta_i^2 x_\mathrm{p}^* \bm{y}_{\mathrm{p},i}}{\eta_i^2|x_\mathrm{p}|^2+\sigma^2},\quad i=1,2,\cdots,n_t,\\
\mathbf{E}[\bm{H}_i\bm{H}_i^*|\mathsf{Y}_\mathrm{p}]&=\frac{\eta_i^4|x_\mathrm{p}|^2 \bm{y}_{\mathrm{p},i} \bm{y}_{\mathrm{p},i}^*}{\left(\eta_i^2|x_\mathrm{p}|^2+\sigma^2\right)^2}+\frac{\eta_i^2\sigma^2}{\eta_i^2|x_\mathrm{p}|^2+\sigma^2}\mathsf{I}_{n_r},
\end{aligned}
\end{equation}
and we deduce that
\begin{equation}
\begin{aligned}
I^{(i)}_{\mathrm{GMI,lin}}\!=\mathbf{E}\bigg[&\log\Big(1\!+\!a_i\bm{Y}_{\mathrm{p},i}^* \big(a_{i+1}\bm{Y}_{\mathrm{p},i+1}\bm{Y}_{\mathrm{p},i+1}^*+\cdots\\
&+a_{n_t}\bm{Y}_{\mathrm{p},n_t}\bm{Y}_{\mathrm{p},n_t}^*+b\mathsf{I}_{n_r}\big)^{-1}\bm{Y}_{\mathrm{p},i}\Big)\!\bigg],
\label{eqn:gmi-lin-case}
\end{aligned}
\end{equation}
where
\begin{equation}
\begin{aligned}
a_i&=\frac{P_i\eta_i^4|x_\mathrm{p}|^2}{\left(\eta_i^2|x_\mathrm{p}|^2+\sigma^2\right)^2},\quad i=1,2,\cdots,n_t,\\
b&=\frac{P_1\eta_1^2\sigma^2}{\eta_1^2|x_\mathrm{p}|^2+\sigma^2}+\cdots+\frac{P_{n_t}\eta_{n_t}^2\sigma^2}{\eta_{n_t}^2|x_\mathrm{p}|^2+\sigma^2}+\sigma^2.
\end{aligned}
\end{equation}

Applying the matrix determinant lemma to \eqref{eqn:gmi-lin-case}, we obtain
\begin{equation}
\begin{aligned}
&I^{\mathrm{sum}}_{\mathrm{GMI,lin}}=\sum_{i=1}^{n_t}I^{(i)}_{\mathrm{GMI,lin}}\\
&=\!\mathbf{E}\!\left[\log\det\!\left(\mathsf{I}_{n_r}\!\!+\!\frac{a_1}{b}\bm{Y}_{\mathrm{p},1}\!\bm{Y}_{\mathrm{p},1}^*\!+\cdots+\!\frac{a_{n_t}}{b}\bm{Y}_{\mathrm{p},n_t}\!\bm{Y}_{\mathrm{p},n_t}^*\right)\right].
\label{eqn:gmi-lin-mimo}
\end{aligned}
\end{equation}

\begin{figure*}[ht]
	\subfigure[$n_t=2$,  $n_r=8$.]{
		\begin{minipage}[t]{0.33\linewidth}
			\centering
			\includegraphics[width=2.3in]{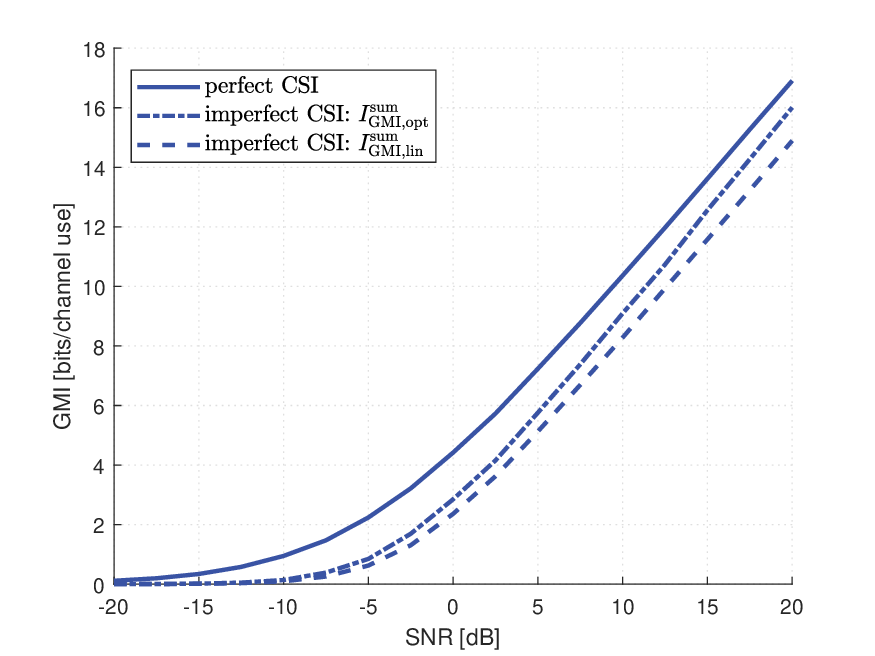}
	\end{minipage}}
	\subfigure[$n_t=4$,  $n_r=8$.]{
		\begin{minipage}[t]{0.33\linewidth}
			\centering
			\includegraphics[width=2.3in]{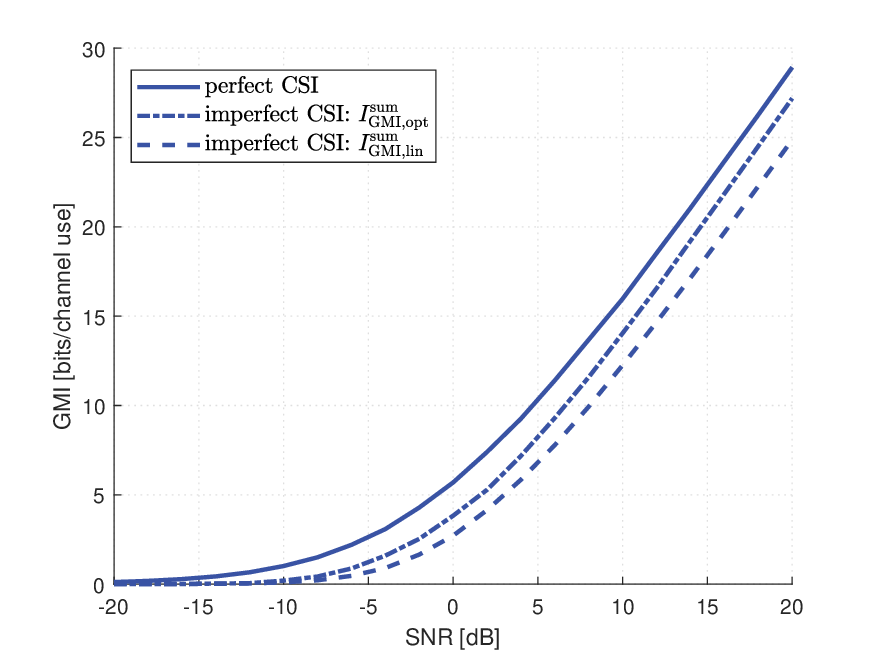}
	\end{minipage}}
	\subfigure[$n_t=8$,  $n_r=8$.]{
		\begin{minipage}[t]{0.33\linewidth}
			\centering
			\includegraphics[width=2.3in]{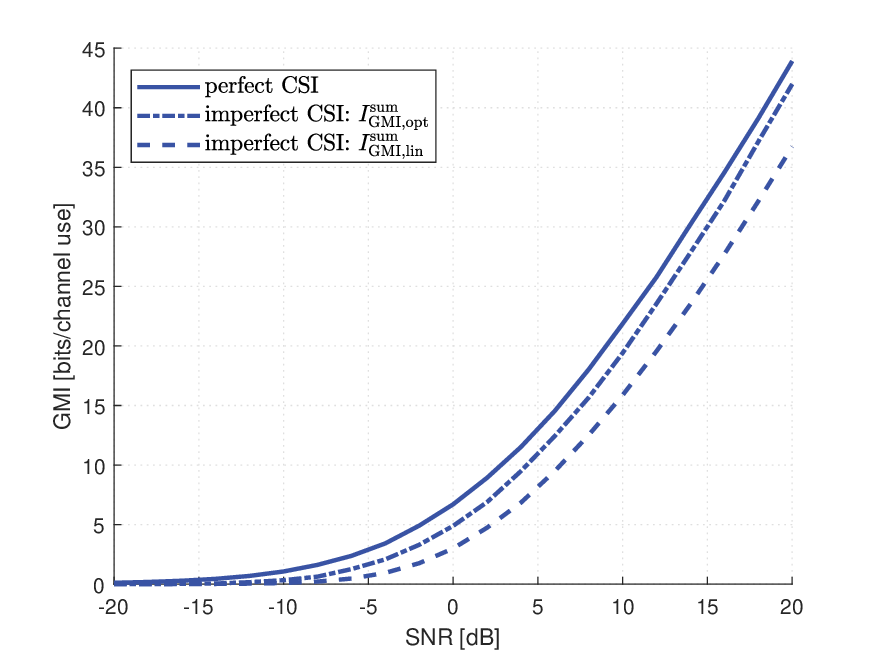}
	\end{minipage}}
	\caption{GMIs for  MIMO fading  channel \eqref{eqn:channel} with perfect and imperfect receiver CSI, $n_r = 8$.}
	\label{fig:nr=8}
\end{figure*}
\begin{figure*}[ht]
	\subfigure[$n_t=2$,  $n_r=16$.]{
		\begin{minipage}[t]{0.33\linewidth}
			\centering
			\includegraphics[width=2.3in]{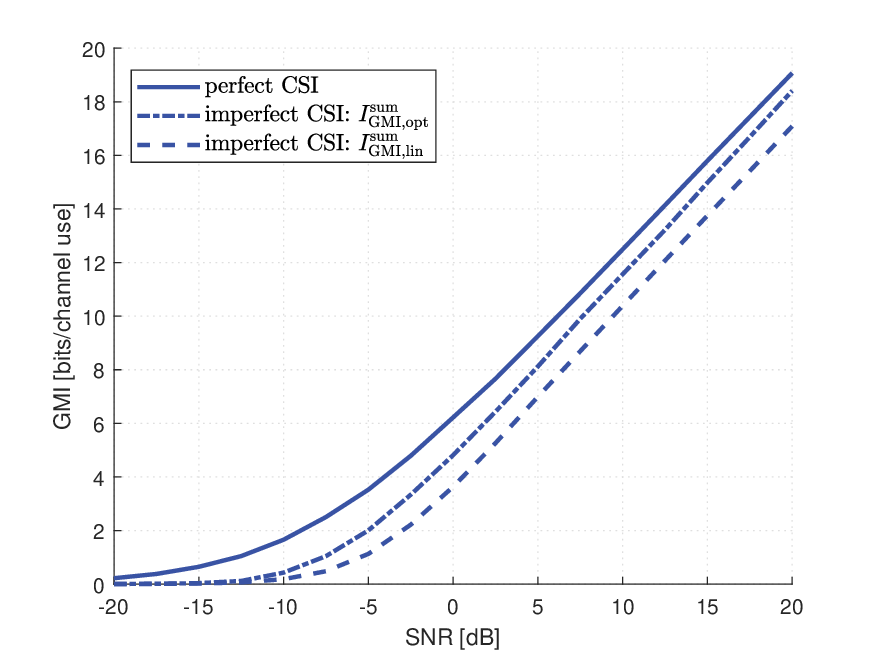}
	\end{minipage}}
	\subfigure[$n_t=4$,  $n_r=16$.]{
		\begin{minipage}[t]{0.33\linewidth}
			\centering
			\includegraphics[width=2.3in]{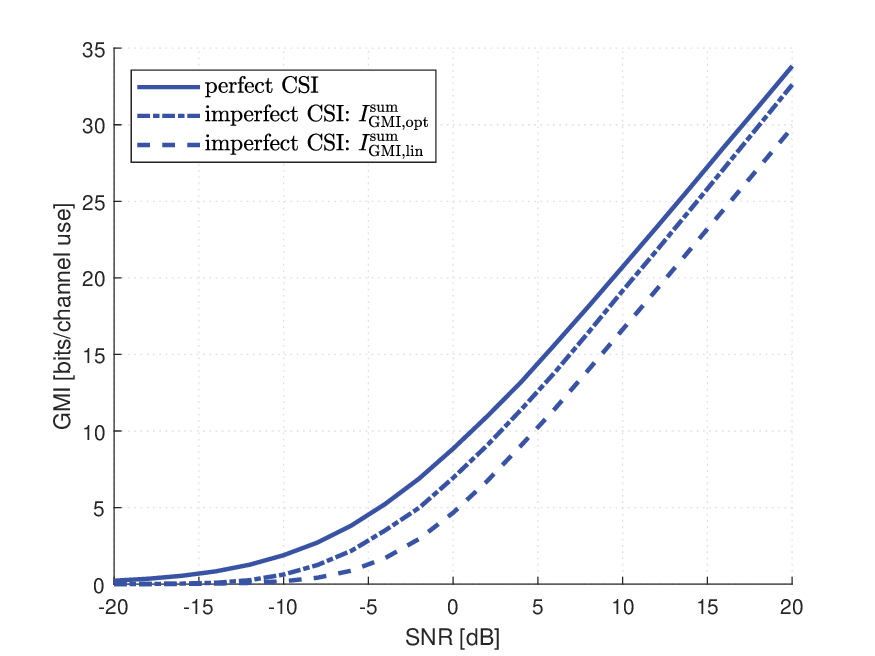}
	\end{minipage}}
	\subfigure[$n_t=8$,  $n_r=16$.]{
		\begin{minipage}[t]{0.33\linewidth}
			\centering
			\includegraphics[width=2.3in]{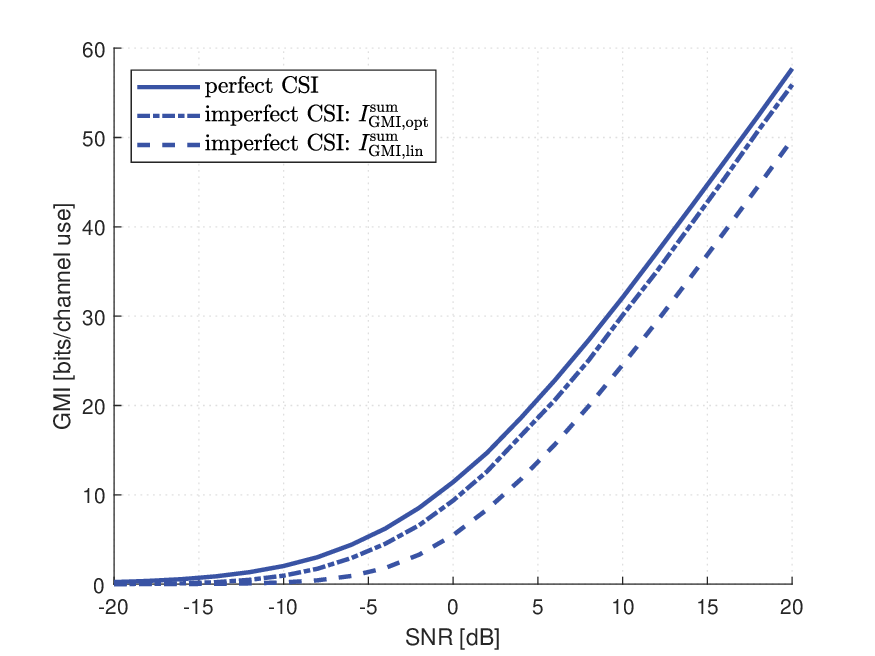}
	\end{minipage}}
	\caption{GMIs for  MIMO fading channel \eqref{eqn:channel} with perfect and imperfect receiver CSI, $n_r = 16$.}
	\label{fig:nr=16}
\end{figure*}
Denoting the MMSE estimate of $\mathbb{H}$ upon observing $\mathbb{Y}_\mathrm{p}$ as $\hat{\mathbb{H}}$, we have
\begin{equation}
\begin{aligned}
\hat{\mathbb{H}}=\mathbf{E}\left[\mathbb{H}|\mathbb{Y}_\mathrm{p}\right]=\left[\frac{\eta_1^2 x_\mathrm{p}^* \bm{Y}_{\mathrm{p},1}}{\eta_1^2|x_\mathrm{p}|^2+\sigma^2},\cdots,\frac{\eta_{n_t}^2 x_\mathrm{p}^* \bm{Y}_{\mathrm{p},n_t}}{\eta_{n_t}^2|x_\mathrm{p}|^2+\sigma^2}\right].
\end{aligned}
\end{equation}

To compare the GMI \eqref{eqn:gmi-lin-mimo} with that achieved by the conventional pilot-assisted scheme, we rewrite the channel model \eqref{eqn:channel} as
\begin{equation}
\bm{Y}=\mathbb{H}\bm{X}+\bm{Z}=\hat{\mathbb{H}}\bm{X}+\underbrace{\left(\tilde{\mathbb{H}}\bm{X}+\bm{Z}\right)}_{\tilde{\bm{Z}}},
\end{equation}
where $\tilde{\bm{Z}}$ is interpreted as the ``overall noise'' of the channel, containing the additive Gaussian noise $\bm{Z}$ and the channel estimation error $\tilde{\mathbb{H}}\bm{X}$. Recognizing that
\begin{equation}
\begin{aligned}
\mathsf{K}_{\tilde z}&=\mathbf{E}\left[\tilde{\mathbb{H}}\mathsf{K}_x\tilde{\mathbb{H}}^*+\bm{Z}\bm{Z}^*|\mathbb{Y}_\mathrm{p}\right]=b\mathsf{I}_{n_r},
\end{aligned}
\end{equation}
we can rewrite \eqref{eqn:gmi-lin-mimo} as
\begin{equation}
\begin{aligned}
I^{\mathrm{sum}}_{\mathrm{GMI,lin}}&=\mathbf{E}\left[\log\det\left(\mathsf{I}_{n_r}+\mathsf{K}_{\tilde z}^{-1}\hat{\mathbb{H}}\mathsf{K}_x\hat{\mathbb{H}}^*\right)\right].
\end{aligned}
\end{equation}
This is exactly the achievable rate derived in \cite{how much} \cite[Thm. 2]{Weingarten}.

Figures \ref{fig:nr=8} and \ref{fig:nr=16} display the GMIs evaluated for MIMO fading channels with 8 and 16 receive antennas, respectively. We let all substreams have equal power $P_i = P/n_t$ for $i = 1,\ldots,n_t$, and let the power of each pilot symbol at a transmit antenna be $P$ so as to satisfy the average power constraint. The solid curves correspond to the GMIs with perfect receiver CSI (also the coherent channel capacity), the dash-dot curves correspond to the maximum GMIs under optimal S-GNNDR (Theorem \ref{thm:mimo-gmi-opt}) with imperfect receiver CSI, and the dashed curves correspond to the GMIs under linear S-GNNDR (Theorem \ref{thm:mimo-gmi-lin}) with imperfect receiver CSI (also the achievable rate of the conventional pilot-assisted scheme \cite{how much} \cite[Thm. 2]{Weingarten}).

We observe that, with imperfect receiver CSI, $I^{\mathrm{sum}}_{\mathrm{GMI,opt}}$ is always larger than $I^{\mathrm{sum}}_{\mathrm{GMI,lin}}$, and their gap increases with the number of transmit antennas; for example, when the SNR is $10$dB and $n_r = 16$, for $n_t=2$, the gap is slightly larger than $1$ bit/channel use or approximately 10\%, whereas for $n_t=8$, the gap becomes nearly $5$ bits/channel use or approximately 20\%. Furthermore, as the number of transmit antennas increases or the SNR increases, $I^{\mathrm{sum}}_{\mathrm{GMI,opt}}$ gets close to the channel capacity with perfect receiver CSI; for example, with $n_t = 8$ and $n_r = 16$, when achieving $40$ bits/channel use, the gap between $I^{\mathrm{sum}}_{\mathrm{GMI,opt}}$ and the channel capacity under perfect receiver CSI is within a fraction of dB, whereas the gap between $I^{\mathrm{sum}}_{\mathrm{GMI,lin}}$ and the channel capacity under perfect receiver CSI is more than 3dB.

\section{Conclusion}
\label{sec:conclusion}

We have proposed a new receiver architecture, termed S-GNNDR, for MIMO fading channels with perfect or imperfect receiver CSI. The basic idea is to integrate a recently proposed GNNDR module into an SIC-like successive procedure, so as to efficiently extract the information embedded in the joint observation of channel output and CSI. Employing GMI as performance measure, we study the achievable rate of S-GNNDR. A restricted linear form of the S-GNNDR is shown to be equivalent to the conventional pilot-assisted scheme, whereas the optimal S-GNNDR yields significant rate gain for channels with imperfect CSI. Future research directions include low-complexity implementation of the optimal S-GNNDR,  its  finite-alphabet  coded modulation design,  and its bit error rate performance.



\end{document}